\newcommand{\avg}[1]{\langle{#1}\rangle}

\newcommand{\req}[1]{(\ref{#1})}

\newcommand{\beq}{\begin{equation}}
\newcommand{\beqar}{\begin{eqnarray}}
\newcommand{\eeqar}{\end{eqnarray}}
\newcommand{\beqars}{\begin{eqnarray*}}
\newcommand{\eeqars}{\end{eqnarray*}}
\newcommand{\eeq}{\end{equation}}
\documentstyle[aps,twocolumn]{revtex}
\begin{document}\twocolumn[\hsize\textwidth\columnwidth\hsize\csname
@twocolumnfalse\endcsname
\title{Dynamical Optimization Theory of a Diversified
Portfolio}
\author{Matteo Marsili$^{(1)}$, Sergei Maslov$^{(2)}$, 
and Yi-Cheng Zhang$^{(1)}$}
\address{$(1)$ Institut de Physique Th\'eorique, 
Universit\'e de Fribourg, CH-1700, Switzerland, \\ 
$(2)$ Department of Physics, 
Brookhaven National Laboratory,
Upton, NY 11973
} 
\date{\today}
\maketitle
\widetext
\begin{abstract}
We propose and study a simple model of dynamical redistribution
of capital in a diversified portfolio. We consider a hypothetical
situation of a portfolio composed of $N$ uncorrelated stocks.
Each stock price follows a multiplicative random walk 
with identical drift and dispersion. 
The rules of our model 
naturally give rise to power law tails in the distribution of 
capital fractions invested in different stocks. The exponent of this 
scale free distribution is calculated in both discrete and continuous
time formalism. It is demonstrated that the dynamical redistribution 
strategy results in a larger typical growth rate of the capital
than a static ``buy-and-hold'' strategy. In the large $N$ limit 
the typical growth rate is
shown to asymptotically approach that of the expectation value of the
stock price . The finite dimensional variant 
of the model is shown to describe the partition function of directed
polymers in random media.
\end{abstract}
]
\narrowtext
\vskip 0.5truecm
\section{Introduction}
The problem of finding an investment strategy with the best
long-term growth rate of the capital is of tremendous 
practical importance.
The traditional theory of portfolio optimization is 
stationary in origin \cite{merton}.
It answers the question of optimal distribution
of the capital between different assets 
(optimal asset allocation), but in general gives no 
prescription on how to maintain this optimal allocation 
at all times. In this work we propose a simple model of 
{\it dynamical} allocation of capital. 
Somewhat counterintuitively,  
in order to optimize the growth rate of the capital
an investor has to sell assets which have increased in 
price since the last update, and buy those which have 
decreased.  In doing so he sells 
stocks when they are ``overpriced'' and buys them 
when they are ``underpriced'', which is clearly 
advantageous.
As we demonstrate below, in our model an investor 
who actively manages his portfolio in such a fashion 
almost certainly does better than one who follows 
a static ``buy-and-hold'' strategy.

%
%
%

The nontrivial properties of the problem come 
from the multiplicative nature of stock 
price fluctuations.
Throughout this manuscript we assume
that on timescales of interest to us the prices of individual
assets follow a multiplicative random walk.
In other words, the ratio of stock prices at 
two consecutive times, at which the investor buys or sells
stock, is a random number, uncorrelated with the current price 
and with the history of price changes in the past.
There are many peculiarities of such noisy multiplicative 
dynamics, especially regarding expectation 
values of random variables. Traditional 
expectation (average) value 
is of little relevance here.
The reason for this is that the dominant contribution to 
the expectation value of a random variable subject to 
multiplicative noise comes from exponentially unlikely outcomes when
the variable is exponentially large. For any finite number of 
realizations (and in real world one always deals with just one 
realization) this expectation (average) value is very unlikely 
to appear. On the other hand, the {\it typical} value of 
such random variable, defined
as the median of its probability distribution, constitutes 
a more realistic property.

Just like in the static portfolio theory, our 
strategy favors the diversification,
i.e. increasing the number of assets in the 
portfolio. We demonstrate that in our model 
the diversification reduces fluctuations, and 
makes the growth rate of the 
typical value of the capital to be closer to that of 
its expectation value. 
However, for any finite number of assets, 
these two growth rates are still 
different.

Under the rules of 
dynamical redistribution of funds, which 
we employ in this manuscript, the distribution 
of shares of 
the total capital invested in individual assets 
naturally acquires a power law tail.
This adds yet another example of how 
a scale free distribution can arise out of
multiplicative dynamics without fine-tuning of any sort. 
We derive the
analytical expression for the exponent $\tau$ of this power law. 
Somewhat surprisingly,  
in the weak coupling limit, corresponding to slow redistribution of
funds between the assets, this exponent has a ``superuniversal'' value
$\tau=2$. It gradually increases with the coupling constant
and becomes infinite in the limit where the capital is 
equally redistributed between assets after each time step. 

The rules of redistribution of capital can be interpreted as
fully-connected (infinite-dimensional) limit of 
the well known statistical model of directed polymers
in the presence of quenched disorder. This provides a new and exciting
link between the physics of finance, and the problems lying 
on the forefront of modern theoretical condensed
matter physics.

The plan of the manuscript is as follows: to streamline the following 
introduction of our basic model, in Section II we review the well known 
(and not so well known) properties of a stochastic multiplicative 
dynamics. We remind the reader the formulas for average and typical 
value of a single  multiplicative random walk, formulate the 
``continuous'' time approach to this problem,  and refresh in reader's 
memory the formalism of Ito stochastic calculus, necessary for our 
purposes. Then we review recent results on natural appearance of power 
law distributions in a situation when a single multiplicative random 
walk is pushed against lower wall \cite{solomon,cont}, preventing the 
random variable from falling below certain value. Finally, we describe 
the multiplicative stock price and capital dynamics used throughout this 
manuscript. 

In Section III we analyze the behavior of the typical and average
values of the capital in a ``buy-and-hold'' 
strategy, where the capital was initially equally 
distributed between $N$ independent assets with the 
same typical growth rate and dispersion, and no 
further redistribution ever took place.
We demonstrate that after a logarithmically short
initial period of time, the typical growth rate of the capital
is limited to the typical growth rate of the price of 
the assets, and is
significantly smaller than their average (expectation) 
growth rate (or average return per capital of this asset)

In Section IV we show that the growth rate of investor's
capital can be significantly increased
by following an active, dynamic redistribution strategy.
In this strategy at each time step the investor sells
some shares of every stock with current value of invested 
capital above the all-stock average, and buys
some shares of every stock below this average.
We analyze the consequences of this strategy in both discrete 
and continuous time formalisms and demonstrate that in both cases these
rules naturally give rise to a scale free distribution of 
fractions of individual stock capitals in the total capital. 
We proceed with deriving analytical  
expressions for the exponent of this distribution, and 
the typical growth rate of the capital in this situation.
This rate for our strategy proves to be larger than that in the static
``buy-and-hold'' strategy, which {\it a posteriori} justifies 
our approach. However, as should be expected, the total capital 
is still subject to the multiplicative noise, and
therefore its typical growth rate is still smaller than the 
average growth rate. We demonstrate that in the limit $N \to \infty$ 
these two rates asymptotically converge as some power of $1/N$.

\section{Review of Results for a Single Multiplicative 
Random Walk}
\subsection{Typical and average values of a multiplicative 
random walk}
Consider a stochastic process in which at each time step  
a variable $W(t)$
is multiplied by a positive random number $e^{\eta (t)}$, where
$\eta (t)$ is drawn from some probability distribution 
$\pi (\eta)$:
\begin{eqnarray}
W(t+1)=e^{\eta (t)} \ W(t) .  
\end{eqnarray}
We adopt the initial condition $W(0)=1$.
For the new variable $h(t)=\ln W(t)$ this process is 
just a random walk with an average drift 
$v=\langle \eta \rangle$ and a dispersion 
$D=\langle \eta ^2 \rangle-\langle \eta \rangle ^2$.
The corresponding equation of motion is simply
\begin{eqnarray}
h(t+1)=h(t) + \eta (t) .
\label{em_w}
\end{eqnarray}

In recent literature it has been observed that 
$average$ and $typical$ values of $W(t)$ in such a
process can be
very different. One of the  precise definitions of the 
typical value of a random variable is the $median$ 
\cite{gnedenko} of its probability distribution, 
i.e.  for $W_{\rm typ}$ one has the property that
${\rm Prob}(W>W_{\rm typ})={\rm Prob}(W<W_{\rm 
typ})=1/2$. By definition $W_{\rm typ}(t) =e^{h_{\rm typ}(t)}$.
  
The central limit theorem implies that asymptotically
the distribution $P(h,t)$ can be approximated with
a Gaussian 
\beq
P(h,t)={1 \over \sqrt{2 \pi D t}} 
\exp(-{(h-v t)^2 \over 2Dt}).
\label{gauss}
\eeq
Therefore, the median (as well as average and  most 
probable values) of $h(t)$
changes linearly with time, and the rate of this change 
is given
by the drift velocity $v=\langle \eta \rangle$
of the corresponding random walk:
$\ln W_{\rm typ}(t)=\langle \ln W(t) \rangle=
\langle \eta \rangle t$.

On the other hand the expectation (average) value of $W(t)$
changes as 
$\langle W(t+1) \rangle=\langle e^{\eta} \rangle \langle 
W(t) \rangle$
(since $\eta(t)$ and $W(t)$ are uncorrelated). Hence,
$\ln \langle W(t) \rangle=\ln \langle e^\eta \rangle \ t$ 
also depends linearly on time but with a different slope. 
It is easy to show that for any distribution 
$\ln \langle e^\eta \rangle > \langle \eta \rangle$, 
so that
the average value of $W$   always grows faster than its 
typical
value and after some time one has 
$\langle W(t) \rangle \gg W_{\rm typ} (t)$. 
This exponentially large  discrepancy between typical  and 
average values of $W$ is due to the long tails of 
$P(W,t)$, but the events constituting these tails are 
extremely rare.

For future use we derive analytic 
expressions for the growth rate of
$\langle W^m (t) \rangle $ in a simple case, when
$\eta$ is drawn from a Gaussian distribution with 
 average value $v=\langle \eta \rangle$
and dispersion $D=\langle \eta ^2 \rangle-\langle \eta 
\rangle ^2$. Since the dynamics of $W^m$ is given by 
$W^m(t+1)=e^{\eta (t) m} W^m(t)$, for $\avg{W^m(t)}$
one has $\avg{W^m(t)}=\avg{e^{\eta m}}^t$.
The integral $\int_{-\infty} ^{\infty} d \eta \ e^{\eta m} 
\ e^{-(\eta-v)^2/ 2D}/\sqrt{2 \pi D}$ \ can be easily 
taken and 
is equal to $e^{ m ( v + D m /2)}$. Therefore, for a 
Gaussian distribution one has
\begin{eqnarray}
\langle e^{\eta m} \rangle ^{1/m} =  e^{v + Dm/2} , 
\label{eta_moments}\\
 \langle W^m(t)\rangle = e^{m(v+ Dm/2)t}.
\label{w_moments}
\end{eqnarray}

It is important to mention that, 
although by the virtue of the Central Limit Theorem,
for any $\pi(\eta)$ with a given average $v$
and dispersion $D$ the distribution $P(h,t)$ can be 
approximated by a Gaussian \req{gauss}, the precision
of this approximation
is not sufficient to calculate averages of the type 
$\langle W^m (t) \rangle = \int e^{m h(t)} \ P(h,t) dh $. 
This integral
is too sensitive to the precise shape of the 
distribution at the upper tail (or lower tail for $m<0$). 
Indeed, the
growth rate of $\ln \langle W^m (t)\rangle$  equal to
$\ln \int_{-\infty} ^{\infty} d \eta \ \pi ({\eta}) 
\ e^{\eta m}$, depends on the whole shape of $\pi (\eta)$
and not only on its first and second moments $v$ and $D$.  

\subsection{Multiplicative random walk in the continuous
time approach}
The above multiplicative process is defined without ambiguity for
discrete time. Straightforwardly taking the continuum limit causes
problems. It might be useful to rewrite the equation of motion of a 
multiplicative random walk as a Stochastic Differential Equation (SDE) 
in continuous time. One should always keep in mind that a stochastic 
differential equation is nothing more than a convenient notation to 
describe a stochastic process in discrete time. At the $n$-th time step 
of discretized dynamics we define a new ``continuous'' time variable $t$ 
as $t=n \Delta t$. Here we introduced a rescaling factor $\Delta t \ll 
1$, which makes one step of underlying discrete dynamics an 
``infinitesimally small'' increment of the continuous time $t$. 
In the SDE approach one is limited to Gaussian distributed
random variables, so we select a Gaussian distribution
of $\eta(t)$ in our discrete dynamics.  
Since we want to approximate $W(t)$ with a continuous 
function, the difference
$W(t+\Delta t)-W(t)$ after one step 
of discrete dynamics should be ``infinitesimally'' small. 
Therefore, we should select both the average value and the 
dispersion of the Gaussian variable $\eta (t)$ to 
scale as some power of $\Delta t$. It turns out 
to be the right choice to make them both 
scale linearly with $\Delta t$: 
$\eta (t)=v \Delta t + \delta \eta (t)$, where
$\avg{\delta \eta (t)^2}=D \Delta t$. Now one can write
the discrete equation of motion for $W(t)$ as
$W(t+\Delta t)=e^{v \Delta t + \delta \eta (t) }\  W(t) \simeq [1+
v\Delta t+ \delta \eta (t)+(v \Delta t + \delta \eta (t))^2/2] W(t)
\simeq W(t) + ((v+D/2) \Delta t +\delta \eta (t)) W(t)$, 
where we have
dropped all terms smaller than linear in $\Delta t \ll 1$.
The SDE for $W(t)$ can now be written as
\begin{equation}
{d W (t) \over d t}=(v+{D \over 2}) W(t) + \tilde {\eta} (t) W(t).
\label{cont_w}
\end{equation}
Here $\tilde {\eta}(t)=\delta \eta (t)/\Delta t$ 
is a usual gaussian ``continuous noise'' with zero mean 
and correlations given by
$\avg{\tilde {\eta} (t) \tilde {\eta} (t')}=D \delta (t-t')$.
We also assume the absence of correlations between $W(t)$
and $\tilde{\eta}(t)$.
This assumption corresponds to selecting
the Ito calculus over Stratonovich calculus.
Both are just two formal ways of linking the 
polemic continuum limit and the well defined
discrete version. 

The nontrivial part of this equation is an extra $D/2$ 
term added to a deterministic growth rate of 
$W(t)$.
This term is not an artifact of our approach, but has a 
real 
physical meaning. Indeed, Eq. (\ref{cont_w}) can be solved
for $\avg{W(t)}$ to give 
$\avg{W(t)} = \avg{W(0)} e^{(v+D/2) t}$, which is the
right answer (see Eq. (\ref{w_moments})). Without this 
extra term we would be lead to the conclusion that for $v=0$, 
i.e. $\avg{\eta}=0$, the average (not typical!) $W(t)$ does
not grow, which is wrong.

The other way to get this extra term in the equation for 
$W$ is to start with the well known   
Langevin equation of motion for $h(t)=\ln W(t)$
describing a usual random walk with a drift:
\begin{equation}
{d h (t) \over d t}= v + \tilde \eta (t) ,
\label{cont_h} 
\end{equation}
where again $\avg {\tilde {\eta} (t)}=0$, and  
$\avg{\tilde {\eta} (t) \tilde {\eta} (t')}=D \delta (t-t')$.
To derive the equation of motion for $W(t)=e^{h(t)}$
one has to do the change of variables as for usual 
partial
differential equations. But in addition to this one has 
to add the ``Ito term'' \cite{ito} given by 
${D \over 2} {\partial^2 W \over \partial h^2}$, which
is a formal prescription of Ito calculus. 
With this nontrivial correction one recovers the 
equation of motion (\ref{cont_w}).
So in the language of SDE the difference between the typical ($v$)
and the average ($v+{D \over 2}$) growth rates of $W(t)$ 
in the multiplicative random walk 
is a direct consequence of the Ito term, appearing
after the change of variables from $h$ to $W$ in the 
equation (\ref{cont_h}).

\subsection{Multiplicative random walk in the presence of a lower wall}
Much attention was devoted recently
\cite{solomon,cont} to the analysis of the
problem of ``multiplicative random walk, repelled
from zero''.
%
In the economical context it was first introduced 
by Solomon {\it et al.} \cite{solomon}.
In a simplest case one has
a multiplicative random walk with 
a Gaussian random variable $\eta$,  
having a  negative average
$v=\langle \eta \rangle < 0$,  
and the dispersion
$D$. In other words the typical value of $W(t)$
exponentially decreasing in time, while its
average may or may not grow in time depending
on the sign of $v+D/2$. 
In addition to this one has
an ``external force'', pushing $W(t)$ up 
and preventing it from falling 
below some predetermined constant. 
This external influence, which will be 
referred to as ``lower wall'', should not
significantly affect the dynamics for large $W$.
One way to introduce a lower wall is to 
add an additional positive  
``source'' term $b$ into the RHS of Eq. \req{cont_w}.
The eqs. \req{cont_w}, and \req{cont_h} now become
\beqar
{dW(t) \over dt} &=& (v+D/2)W(t)+\eta(t)W(t)+b ; 
\label{cont_w_wall}\\
{dh(t)\over dt}&=&v+\eta(t)+b \exp(-h).
\label{cont_h_wall}
\eeqar 
As we see the lower wall in Eq. \req{cont_h_wall} 
has a property of
being ``short-ranged'' in $h$-space, i.e. its contribution
to the SDE for $h(t)$
can be neglected for large positive $h$. But for negative $h$
the strength of the wall grows exponentially and compensates
the negative drift already at $h=-\ln(|v|/b)$.
It is easy to convince oneself that this 
stochastic process
eventually reaches a stationary state, characterized by a
stationary probability distribution $P(h)$.
In this stationary state the negative drift of
$h(t)$ is precisely balanced with diffusion 
combined with repulsion from the lower wall.

In the literature on this subject one encounters
many different realizations of the lower wall mechanism.
For instance, one can introduce a more general  
term $b W^{\delta}$ 
into the RHS of  \req{cont_w} \cite{grinstein,hwa}. 
In the equation for $h$ this term 
becomes $b e^{(\delta -1)h}$, which for any $\delta<1$
describes an exponential lower wall qualitatively
similar to \req{cont_h_wall}. Indeed, the ``source'' term 
in \req{cont_w_wall}
is just a particular example of this more general term 
with $\delta=0$.
On the other hand, the terms with $b<0$ and $\delta>1$
describe an ``upper wall'', preventing $h$ from
becoming to big. In this case, in order for a 
stationary state to exist one needs a positive drift of $h$
pushing it up against the wall. In \cite{solomon}
the lower wall is introduced ``by hand'': in their 
simulations the authors simply do not allow $h(t)$ to 
fall below a predetermined constant $h_{min}$. 
In other words,  $h(t+1)=\min (h(t)+\eta(t), h_{min})$.
Such ``infinitely hard lower wall'' can be described by 
a term $b W^{\delta}$ with very large 
{\it negative} $\delta$. Finally, Cont and Sornette \cite{cont}
consider a case when the constant $b$ 
itself can depend on time obeying a deterministic
and/or stochastic dynamics. Except for pathological cases, 
where typical $b(t)$ exponentially grows or decays
in time, it does not qualitatively change the results, 
compared to a time-independent lower wall 
\cite{solomon}.

An interesting feature of a 
multiplicative random walk with a lower 
wall is that it generically gives rise to a power law 
tail in the distribution of $W$ in the stationary state.
We proceed by reviewing various derivations of this  
result found in recent literature \cite{solomon,cont}.
As was explained above, the lower wall's only purpose is to 
make the process stationary by pushing the variable
up whenever it becomes too small. 
The drift due to the wall can always be neglected
for large enough $h$. 
In the region, where 
this approximation is justified
one can write a Fokker-Planck equation, 
taking into account only the  
multiplicative part of the process, 
equivalent to diffusion with a drift in the $h$-space.
The stationary solution of the Fokker-Planck
equation should satisfy 
$-v {\partial P(h) \over \partial h}+{D \over 2} 
{\partial^2 P(h) \over \partial h ^2}=0$.
It is easy to see that $P(h)=A\ \exp (2v h /D)$ 
is indeed a solution. Since $v<0$, it 
exponentially decays for 
positive $h$. 
The deviations from this form start to appear 
only at low $h$, where the presence of lower wall 
cannot be neglected. This ``Boltzmann'' tail of the 
distribution of $h$ corresponds to a power law tail 
of distribution of $W=e^h$: $P(W)=AW^{-1+2v/D}$.
The exponent of this power law tail
\beq
\tau=1-2v/D=1+2|v|/D 
\label{tau_cont}
\eeq
is greater than 1, so that there are 
no problems with normalization. In case 
of a lower wall of the form $b e^{-h}$ 
(see Eq. \req{cont_h_wall})
one can write an analytic 
solution of the Fokker-Planck equation valid for
any $h$. It is the Boltzmann distribution
with a Hamiltonian $H(h)=b e^{-h}-v h$ and
temperature $T=D/2$, i.e. 
$P(h)= A \exp[(-2b e^{-h}/D +2v h / D)]$, 
or $P(W)=A\exp ( -2b/DW)W^{-1+2v/D}$. 
The normalization constant $A$ is given by
$A=(2b/D)^{-2v/D}/\Gamma (-2v/D)$.

The Eq. \req{tau_cont},  expressing the exponent
of the power law tail of $P(W)$ in terms of $v$ and $D$, 
is valid only for the case of Gaussian distribution $\pi(\eta)$.
Indeed, in its derivation we employed a stochastic 
differential equation approach, which is restricted 
to Gaussian noise. It is instructive to
derive an equation, giving the value of $\tau$ 
for a general $\pi(\eta)$.
It was first done by Kesten in \cite{kesten}
and recently brought to the attention of physics
community in \cite{cont}. Again, the formula holds
for any multiplicative process with a negative average
drift ($\avg{\eta}<0$) and a lower wall, the effect of which
can be neglected for large $W$. We assume that the process
has already reached a stationary state, characterized by 
a stationary distribution $P(W)$. For sufficiently large
$W$, so that one can neglect the effect of the wall, 
the stationarity imposes the following functional equation
on $P(W)$: 
$P(W)=\int_{-\infty}^{+\infty} \pi(\eta)\ d\eta
\int_0^{+\infty}P(W')\ \delta(W-e^{\eta}W') dW'=
\int_{-\infty}^{+\infty} d\eta \pi(\eta)
e^{-\eta} P(We^{-\eta})$. Assuming that the solution
has a power law tail $P(W) \sim W^{-\tau}$ one finds
$\int_{-\infty}^{+\infty} d\eta \pi(\eta)
e^{\eta(\tau-1)}=1$. In other words $\tau$ is given by 
a solution of
\beq
\avg{e^{\eta(\tau-1)}}=1.
\label{eq_tau}
\eeq
The obvious solution $\tau=1$ should be rejected because
the distribution function is not normalizable in this case.
In short, we are looking for a solution with $\tau>1$.
Let us define 
$\Lambda (\tau)=\avg{e^{\eta(\tau-1)}}$. 
Since $d \Lambda (1)/d \tau=\avg{\eta}<0$, but
$d^2 \Lambda (\tau)/d\tau ^2 >0$ one has at most
one such a solution. In fact, if the distribution of $p(\eta)$
is not restricted to $\eta<0$, for $\tau \to +\infty$
one has $\Lambda (\tau) \to +\infty$ and the solution
is guaranteed by the continuity of $\Lambda (t)$. 
Only in the situation when $\eta$ is always 
negative, the
region of large $W$ is absolutely inaccessible, 
and no power law tail at large $W$ is feasible.
Using Eq. \req{eta_moments}, 
one can check that for a Gaussian distribution 
Eq. \req{eq_tau} predicts $\tau=1-2v/D$ in agreement with 
\req{tau_cont}.

\subsection{Interpretation of $W(t)$ as a fluctuating 
stock capital}
In what follows we will stick to the following 
``realization'' of the random multiplicative process: 
we interpret $W(t)$ as the capital 
(or wealth, hence the notation) that a 
single investor has in some stock. The price of the share of 
this stock $p(t)$ undergoes 
a random multiplicative process $p(t+1)=e^{\eta (t)}p(t)$, 
and if the investor keeps a fixed number $K$ 
of shares without selling
or buying this stock, his capital $W(t)=K p(t)$ 
follows these price fluctuations.
Later on we will consider models, where the investor
at each time step will sell some stock and buy another.
We assume that volumes of such transactions 
are sufficiently
small, so that they have no influence on 
the market price fluctuations. 
Hence our assumption that $\eta (t)$ and 
$W(t)$ are uncorrelated.

The lesson one derives from the above properties of 
multiplicative 
random walk is that  if the investor keeps all his 
money in just one stock 
it is the {\em typical} growth rate $\langle \eta \rangle$, 
he should be concerned about.
In majority of realizations his
capital grows at {\it typical} rate
and he cannot directly take advantage of a bigger
{\em average} growth rate $\ln \langle 
e^\eta \rangle$. There are situations
when the typical growth rate is negative, i.e. the stock price
is going down, while the fluctuations are strong enough to make
the average rate positive.
The question we are going to address in 
this manuscript is how one can still 
exploit this {\it average} growth rate 
by investing and actively managing a portfolio
composed of $N$ stocks.

\section{Ensemble of $N$ stocks without redistribution.} 
 
The first problem we are going to consider is:  
what is the {\em typical} growth rate of the capital 
invested in an ensemble of $N$ stocks if one is not 
allowed to sell one of them and reinvest the money into another. 
In the following we assume that the price $p_i (t)$ of a share
of each stock undergoes a multiplicative random walk, 
independent of price fluctuations of other stocks. 
In other words, one time step logarithmic price increments
$\eta _i (t)$ are uncorrelated not only at 
different times, but 
also for different stocks at a given time. 
The validity of this approach for the real stock market 
lies beyond the scope of this work.
For simplicity of final expression in this section we will 
restrict ourselves to the situation 
when $\eta _i$ for each of the stocks are Gaussian 
variables with zero
mean ($\langle \eta_i \rangle =v=0$) 
and the same dispersion $D=\langle \eta _i \rangle$. 
Initially the capital is equally distributed between 
all stocks. We assume that the starting capital
in each stock is equal to $1/N$, so that the total 
capital
is equal to $1$. The typical value of the total capital 
$(W_{tot} (t))_{\rm typ} 
=(\sum _{i=1}^N W_i (t))_{\rm typ}$ will then grow in 
time. 
From the results of the previous section one concludes 
that
$\langle W_i(t) \rangle = e^{Dt/2}$ and 
$\langle W_i(t) ^2 \rangle - \langle W_i (t) \rangle ^2 
= e^{2Dt}-e^{Dt}$.
One can safely replace the sum of N variables with their 
average
as long as $((\langle W_i (t) ^2 \rangle - \langle 
W_i(t) \rangle ^2)/N)^{1/2}
\ll \langle W_i (t) \rangle$. Therefore, at short times, 
when
$Dt \ll \ln N$, one indeed enjoys the average growth 
rate:
$(W_{tot} (t))_{\rm typ} =e^{D t/2}$. At later times, 
however,
the typical value of the capital starts to fall below 
the average
value (i.e. average value over infinitely many 
realizations). 
To determine this slower growth of typical value 
quantitatively 
one has to approach the problem
from a different end. At late times the value of the 
total capital is 
mainly determined by the capital accumulated in the most 
successful stock, 
i.e. $W_{\rm tot} (t) \simeq W_{\max}\equiv\max _{i=1,N} 
W_i (t)$. 
The extremal statistics theory \cite{galambos} readily 
gives the 
typical value of the $W_{\max}$
by requiring that $1/N={\rm Prob}(W>W_{\max}) 
={\rm Prob}(\ln W>\ln W_{\max}) 
\sim \exp(- \ln ^2 W_{\max}/2Dt)$ 
With exponential precision one gets $W_{\max} \sim e^{\sqrt{ 2 D t \ln 
N}}$. Our approximation that  $W_{\rm tot} (t) \simeq 
=\max _{i=1,N} W_i (t)$ is good only if the second maximal $W$ (we denote it 
as $W_{\max} ^{(2)} (t)$) is much smaller than the maximal one. 
Following the same arguments as before one concludes to find the typical 
value of $W_{\max} 
^{(2)} (t)$ one needs to solve ${\rm Prob}(W>W_{\max}^{(2)}) = 2/N$. 
This results in $W_{\max}^{(2)} \sim e^{\sqrt{ 2 D t (\ln N-2)}}$. One 
easily confirms that the approximation of the whole sum with its biggest 
element makes sense if $Dt \gg \ln N$, which is a complementary 
condition to the ``average'' growth at small times. Therefore, we 
conclude that 
\begin{eqnarray}
(W_{\rm tot} (t))_{\rm typ} =e^{D t/2} \quad {\rm for 
\quad } t \ll \ln N / D ; \\
(W_{\rm tot} (t))_{\rm typ} = e^{ \sqrt{ 2 D t \ln N}}
\quad {\rm for \quad } t \gg \ln N / D . 
\end{eqnarray}
Since growth proportional to 
$\sqrt{t}$ is slower than linear in $t$ one concludes that no matter how 
big is your $N$ your asymptotic growth of your total capital is still 
determined by the ``typical'' growth rate $v=\langle \eta \rangle$ 
(equal to zero in the case considered above) of a single stock.  

If one wants to exploit the ``average'' growth rate for a 
period of
time $T$ and then sell the stocks one needs to take an 
exponentially 
large ensemble of stocks $N> e^{DT}$.

\section{Ensemble of $N$ stocks with redistribution.}

The case of ``non-interacting'' stocks, considered in the  
previous section, can be also called the case of a ``lazy 
investor''. Indeed, initially 
the investor puts equal capital in $N$ stocks and leaves 
them as they are.
He never sells or buys stocks. No wonder that very 
soon he can no longer expect to get an {\it average} 
rate of return on his investment and has to settle for
smaller {\em typical} growth rate.
Now we are going to consider the case of 
an {\it active} investor who after each 
time step redistributes his capital 
between stocks according to some simple 
rule. One may naively think that by selling unsuccessful 
stocks with
small $W_i$ and reinvesting the money into successful 
stocks with
large $W_i$ one may do better. In reality the answer is 
precisely the opposite: one needs to sell some of the most 
successful stocks and reinvest 
the money into the least successful stocks.
Selling only small number of shares of the most successful stocks
(i.e. ones which are currently overpriced) and reinvesting
this money into the least successful stocks (i.e. ones which
are currently underpriced) makes a 
huge difference: $\ln W_i$ for underpriced stocks goes up 
significantly, while $\ln W_i$ for overpriced stocks does 
not go down as much.
As we will show such a ``charity'' between stocks 
bootstraps the typical growth rate of the capital, so that
$\ln (W_{\rm tot} (t))_{\rm typ}$ at all times 
has a growth rate bigger than a {\it typical} growth rate 
of a single stock. For large $N$ this rate 
quickly approaches the {\it average} growth rate 
given by $\ln \langle e^\eta \rangle $ (equal to $D/2$ 
for the Gaussian
distribution of $\eta$ with zero mean). This growth rate 
serves as a theoretical maximum of all possible 
growth rates achievable by simple redistribution of 
funds.

\subsection{Problem with redistribution in the discrete time approach}

We start with the simplest strategy for 
redistribution of the capital.
Under this strategy at each time step the investor 
calculates the current value of average
capital per one stock $\overline{W(t)}={1 \over N} 
\sum_{i=1,N} W_i(t)$.
The capital is redistributed between the stocks 
according
to the rule $W_i \to W_i-\lambda (W_i -  \overline{W})$. 
For positive $\lambda$ it means that ``overpriced'' 
stocks with $W_i(t) > \overline{W(t)}$ loose
a fraction of their capital in favor of the ``underpriced'' ones 
with $W_i (t) < \overline{W(t)}$. The extremal case of $\lambda=1$
corresponds to the equal redistribution of the capital
after each time step. The stock price changes 
during the next discrete time interval. 
As a result the capital invested in each stock 
is multiplied by the random factor 
$e^{\eta_i (t)}$. 
The complete change of each stock's capital after one time 
step is given by: 
\begin{equation}
W_i(t+1)= e^{\eta_i (t)}[(1-\lambda) W_i (t) + 
\lambda \overline{W(t)}] .
\label{dyn}
\end{equation}

One can recognize the above model can be interpreted as the Directed 
Polymer model in $N$ dimensions, with mean field (fully connected) 
interactions\cite{hz}. The role Laplacian is played by $\overline 
{W(t)}-W_i(t)=(1/N \ \sum_{j=1,N} W_j(t)) - W_i(t)$. It is convenient to 
introduce a new set of rescaled variables $s_i(t) = W_i (t) /  
\overline{W(t)}$. The sum of $s_i$ is always equal to $N$, which sets a 
theoretical cutoff equal to $N$ to a  value of individual $s_i$. One can 
rewrite Eqs. (\ref{dyn}) in the following form: 
\begin{eqnarray}
s_i(t+1)= {\overline{W(t)} \over \overline{W(t+1)}}
e^{\eta_i (t)} [(1-\lambda) s_i (t) + 
\lambda] ; \label{dyn_s} \\
\overline{W(t+1)}=\overline{W(t)}{\sum _{i=1,N} 
e^{\eta_i (t)} 
[(1-\lambda) s_i + \lambda] \over N} \label{dyn_bar} .
\end{eqnarray}
As we will confirm later, the dynamics of 
$\overline{W(t)}$
can be approximated as a random multiplicative process, 
where
the multiplication factor $\Gamma (t)=\sum _{i=1,N} 
e^{\eta_i (t)} ((1-\lambda) s_i + \lambda)/N$ has only small 
fluctuations
around its average value. We will indeed demonstrate 
that 
$\Gamma (t) = \langle \Gamma \rangle + \delta \Gamma (t)$, 
where $|\delta \Gamma (t)| \sim N^{-\alpha/2}$. 
It means that for large $N$ to a good 
approximation one can disregard the fluctuations of  
$\overline{W(t+1)}/\overline{W(t)}$ while trying to 
solve Eq. (\ref{dyn_s}).
The average value of this ratio is easily calculated and 
is equal to $\langle e^{\eta}\rangle$ 
(one has to recall that $\sum_{i=1,N} s_i=N$).
In this approximation the equations of motion for $s_i$ 
decouple and allow for exact solution. These {\it  mean-field} 
equations are:
\begin{equation}
s_i(t+1)= {e^{\eta_i (t)} \over \langle e^{\eta}\rangle} 
[(1-\lambda) s_i (t) + \lambda] .
\label{dyn_s_mf}
\end{equation}
Similar equation of motions were recently studied by 
Cont {\it et al.} \cite{cont} and Solomon {\it et al.} 
\cite{solomon}
and were shown to give rise to a stationary distribution 
of $s$ having a power law tail for large $s$. One 
has to keep in mind that the definition of $s$ in our 
problem introduces a natural cut off to 
this tail as $s \leq N$, so it is only for large $N$ 
that one has a chance to
see the effect of this power law or measure this power 
law numerically. 

The stationary distribution $P(s)$ is conserved 
by dynamics. Therefore, it should satisfy the 
following functional equation: 
\begin{equation}
 P(s)=\int d \eta \ \pi (\eta) P({s \over R(\eta)}-
{\lambda \over 1-\lambda})/R(\eta) ,
\label{fe}
\end{equation}
where $R(\eta)=(1-\lambda) e^{\eta}/\langle e^{\eta} 
\rangle$.
Using this equation one can easily verify 
that indeed
$\langle s \rangle = \int s P(s) ds = 1$, which is to be expected
since $\sum_{i=1,N} s_i =N$. Assuming that $P(s)$ has a power 
law tail 
of the form $A s^{-\tau}$, and substituting it to the 
functional equation
(\ref{fe}) one gets the self consistency condition for 
the exponent $\tau$:
$\int d \eta \pi (\eta) R (\eta )^{\tau-1} = 1$, or
\begin{equation}
{\langle e^{\eta (\tau-1)} \rangle ^{1 \over \tau -1 } 
 \over \langle e^{\eta} \rangle }={1 \over 1-\lambda}
\end{equation}
For a general distribution $\pi(\eta)$ this equation 
cannot be solved 
analytically. All one can deduce is that for a weak 
coupling
$\lambda \ll 1$ the solution exists and is approximately 
given by
$\tau=2$. That means that for a weak coupling one always 
has 
$P(s) \sim 1/s^2$ !
For a case of Gaussian distribution of $\eta$ the 
analytic expression
for $\tau$ can be easily obtained from Eq. 
(\ref{eta_moments})
and is given by
\begin{equation}
\tau=2-{2\ln (1-\lambda) \over D} \qquad .
\label{tau}
\end{equation}
In Fig. 1 we present the results of simulations of the model with 
$N=10000$. The measured power law exponent is in excellent agreement with 
the above theoretical prediction.

Our ultimate goal is to determine $\overline{W(t)}_{\rm typ}$ as a 
function of $t$. The Eq. (\ref{dyn_bar}) states that at each time step  
$\overline{W(t)}$ is multiplied by $\Gamma(t)=\sum _{i=1,N} e^{\eta_i 
(t)} ((1-\lambda) s_i +\lambda)/N$. One can show that  $\Gamma(t)$ at 
different time steps are uncorrelated. One can also disregard possible 
correlations between the value of $\overline{W(t)}$ and $\Gamma(t)$ at 
the same time step. Then the behavior of $\overline{W(t)}$ is nothing 
else but a  multiplicative random walk studied in Section 1. The typical 
value of $\overline{W(t)}$  grows as $(\overline{W(t)})_{\rm typ}=e^{t 
\langle \ln \Gamma \rangle}$, while its average value grows as $e^{t \ln 
\langle \Gamma \rangle } = e^{t \ln \langle e^{ \eta }\rangle 
}=\avg{e^{eta}}^t$. 

We will proceed by demonstrating that for any $\lambda >0$ 
the typical and average growth rates of $W(t)$ differ by 
$O(N^{-\alpha})$.
For $\langle \ln \Gamma \rangle$ one has the exact 
expression:
\beq
\langle \ln \Gamma \rangle=\ln \avg{e^{\eta}} \left\langle
\ln \left(1+\frac{1}{N}\sum_{i=1}^N \chi_i [(1-\lambda)s_i+\lambda] 
\right) \right\rangle
\label{gamma}
\eeq
where we introduced the notation
$\chi_i=e^{\eta_i}/\avg{e^{\eta}}-1$.
Expanding the second logarithm for large $N$, 
we get to leading order:
\[
\langle \ln \Gamma \rangle \simeq \ln \avg{e^{\eta}}-
\frac{1}{2N^2}\sum_{i=1}^N
\avg{\chi_i^2} \avg{[(1-\lambda) s_i +\lambda]^2},
\]
where we used the fact 
that $\chi_i (t)$ are uncorrelated at different $i$'s.
Therefore, $\avg{\chi_i(t)\chi_j(t')}=\tilde D
\delta_{i,j}\delta_{t,t'}$, where 
$\tilde{D}=\avg{e^{2\eta}}/\avg{e^{\eta}}^2-1$. 
The fact that these variables are uncorrelated at 
different times proves that
indeed $\overline{W(t)}$ undergoes a multiplicative 
random walk.
The last step is to estimate $\sum _{i=1,N} s_i^2$. To 
do this we need
to recall our results for the stationary distribution 
$P(s)$. If the 
exponent $\tau$ of the power law tail of this 
distribution is larger 
than 3, $\langle s^2 \rangle$ is finite, $\sum _{i=1,N} 
s_i^2= N \langle s^2 \rangle$ and one immediately gets
$\langle \ln \Gamma \rangle=\ln \avg{\Gamma}-A/N$, where 
$A=[(1-\lambda)^2 
\langle s^2 \rangle + \lambda ^2 +2(1-\lambda)]
\tilde D $. In reality 
this is not hundred percent true. Indeed, expanding the 
logarithm in Eq. \req{gamma} we stopped at the first 
order. In the
presence of power law tails in $P(s)$ the validity of 
this
approximation is in doubt because 
the higher order terms involve the sum of powers $s_i^k$ 
with
$k>2$. For large enough $k$ such powers are known to 
diverge as some power of $N$. It can be shown that 
for {\it very large} $N$ they would dominate the 
scaling with respect to $N$. 
Such crossover was indeed observed in simulations.
In Fig. 2 we present the results of the simulations 
of our model with $\lambda=0.1$, $D=0.1$, which corresponds to
$\tau=4.1$. Indeed, we observe that for $N \to \infty$, the
difference between the average and typical growth rates
of the total capital, $v_{avg}-v_{typ} (N)= \ln \avg{\Gamma}-\avg{\ln \Gamma}$, approaches zero.
This approach starts as $A/N^{\alpha}$ with 
$\alpha=1$, but at larger $N$ a 
deviation towards smaller $\alpha$ can be noticed. 

For  $2<\tau<3$ the second moment of $s$  diverges.
This  means that
one should be more careful in estimating $\sum _{i=1,N} 
s_i^2$. 
The apparent divergence of the integral $\int s^2 \ P(s) \ ds $ 
should not be taken too seriously, 
since we are dealing with a finite 
sample of variables $s$ restricted by $\sum s_i=N$.
Even in the worst case if only one $s_i$ is nonzero (and 
equal to N by normalization) the sum $\sum _{i=1,N} 
s_i^2=N^2$.
In all situations when the integral $\int s^k \ P(s) \ ds $ 
diverges the sum of a finite sample is dominated by the 
largest element. 
One can estimate this
largest $s$ by requiring ${\rm Prob} (s>s_{max}) 
=s_{max}^{1-\tau}=1/N$.
Therefore, the typical value of the largest $s_i$ is 
given by
$s_{max} \sim N^{1/(\tau-1)}$. Since $\tau \geq 2$ this 
value is always 
less then $N$ - the maximal possible $s$. 
Then  $\sum _{i=1,N} s_i^2 \simeq  s_{max}^2 \sim 
N^{2/(\tau-1)}$.
Now the expression for the  $\langle \ln \Gamma \rangle$ 
becomes
$\langle \ln \Gamma \rangle=\ln \avg{\Gamma}-A'/N^{-2+{2 \over 
\tau-1}}$, with $A' \sim (1-\lambda)^2 \tilde D$.

\subsection{Problem with redistribution in continuous time approach} 
 
Similar results can be obtained in the continuous
time limit of Eq. \req{dyn}. In order to derive the
stochastic partial differential equation corresponding 
to Eq. \req{dyn} we assume that time is discretized 
$t=n \Delta t$ in units $\Delta t$ and we take $\lambda =\lambda_c \Delta t$, 
$v=v_c \Delta t$, and $D=D_c \Delta t$. 
In all our future formulas we drop the
subscript $c$ in $\lambda_c$, $v_c$, and $D_c$
of continuous model. However, one should keep in mind that
we recover continuous limit by making parameters $\lambda$, $v$, and
$D$ of a discrete model very small, keeping their ratio fixed.
 
In the limit $\Delta t\ll 1$ the Eq. \req{dyn} becomes a stochastic 
differential equation 
\beq
\partial_t W_i(t)=\lambda (\overline W-W_i)+ (v+D/2) \ W_i+W_i \tilde
\eta_i (t).
\label{dync}
\eeq
Here as in Section 1 we introduced the continuous-time stochastic 
force $ \tilde \eta_i(t)=\eta_i(t)/\Delta t -v$,
and used $e^{\eta_i}=1+\eta_i+\eta_i^2/2+\ldots\simeq
1+\tilde \eta _i \Delta t+ (v+D/2) \Delta t+O( \Delta t^{3/2})$.
It is important to point out here that such a continuous
time formulation is only meaningful if $\eta_i(t)$ is
a Gaussian noise. Only in this case Eq. \req{dync} 
can be regarded as a Langevin equation\cite{ito}. Usually
the assumption of a Gaussian noise is motivated by
the fact that for a continuous time 
process, the stochastic force $\tilde{\eta}_i dt$ acting on a
small interval $\Delta t$ can be thought of a sum of 
infinitely many infinitesimal contributions. The 
central limit theorem then ensures that $\tilde{\eta}_i \Delta t$
is Gaussian. For processes with additive noise, 
this assumption is reasonable also for discrete time 
processes. For multiplicative processes the deviations
from the central limit theorem becomes of concern since
the tails of the distributions are probed by the 
process. 
Therefore, we shall assume in this section, 
that $\tilde{\eta}_i$ is Gaussian.

Under this assumption, we shall be able to derive the 
full probability distribution of the $W_i$ in the limit
$N\to\infty$. It is again convenient to use the 
variables 
$s_i(t)=W_i(t)/\overline W(t)$. Using Ito calculus, one 
readily finds 
\beq
\partial_t s_i=\lambda (1-s_i) 
-\frac{D}{N}s_i\left(s_i-\overline{s^2}\right)
+s_i\tilde{\eta}_i -\overline{s \tilde{\eta}}\, s_i
\label{dyns}
\eeq
where we used the notation $\overline 
f=\frac{1}{N}\sum_j f_j$.
Note that $\sum_i s_i= N$ and, consistently, 
$\sum_i \partial_t s_i =0$.
We shall adopt a self consistent 
mean field approach, valid in the 
$N\to\infty$ limit, in which we substitute averages
over $i$ with statistical averages: $\overline f\cong 
\avg{f}$.
Within this approximation, 
the term $\overline{s \tilde{\eta}}\cong \avg{s\tilde{\eta}}=0$
can be neglected. 
If we introduce
\beq
\tau=\frac{2\lambda}{D}-\frac{2}{N}\avg{s^2}\cong 
\frac{2\lambda}{D}-\frac{2}{N}\overline{s^2}
\label{tau1}
\eeq 
as a constant to be determined later 
self--consistently, Eq. \req{dyns} becomes
an equation for $s_i$ only, which does not involve 
$s_j$ for $j\neq i$ explicitly.
We know\cite{ito} that, for a Langevin equation of 
the form
\[
\partial_t s=-s^2\frac{dV(s)}{ds}+s\tilde{\eta},
\]
the associated Fokker Planck equation yields
the asymptotic distribution $P(s)\sim e^{-2V(s)/D}$.
Recasting Eq. \req{dyns} into this form, we find
$V(s)=\lambda/s+{\tau D \ln s \over 2}+D s/N$, 
from which
\beq
P(s)={\cal N}\exp\left[-\frac{2\lambda}{Ds}-
\frac{2}{N}s\right]s^{-\tau}.
\label{pdis}
\eeq
Note the emergence of a power law behavior in $P(s)$,
which is however cut off by the second term in the
exponential. This is physically meaningful, since
$s\le N$ must hold, with $s=N$ occurring when 
the whole capital $N\overline W$ is invested in a 
single stock. The value $\tau$ of the power 
law decay is determined self--consistently from
Eq. \req{tau1} performing the average on the
distribution in Eq. \req{pdis}. A further 
requirement which our approach imposes on 
$P(s)$ is that $\overline s\cong\avg{s}=1$.
It is not possible to compute exactly
these averages, however, it is possible to 
perform a large $N$ expansion. Indeed if we
set
\[
Z(\mu)=\int_0^\infty ds \exp\left[-\frac{2\lambda}{Ds}-
\mu s\right]s^{-\tau}
\]
then, clearly,  $\avg{s}=-\partial_\mu \ln Z(\mu)|_{\mu=2/N}$ and 
$\avg{s^2}=\partial_\mu^2 \ln Z(\mu)|_{\mu=2/N}+\avg{s}^2$. Therefore,  
evaluating the small $\mu$ expansion of $Z(\mu)$ we can compute the 
first two moments of $s$ and impose self--consistency. However $Z(\mu)$ 
has a non--analytic expansion around $\mu=0$, since derivatives 
$\partial_\mu^n Z(\mu)$ diverge at $\mu=0$ for $n\ge \tau$. For 
$\lambda>D/2$, the first two derivatives exist. The equation $\avg{s}=1$ 
then allows us to compute $\tau$ together with its leading correction: 
\beq
\tau\cong 2+\frac{2\lambda}{D}-
\frac{\lambda((D/2)^2-\lambda D/2+\lambda^2)}{(D/2)^2(D/2-\lambda)}
\frac{4}{N}, ~~~{\rm for} D/2<\lambda
\label{tau2}
\eeq
The equation \req{tau1} then turns out to be
automatically satisfied, which is a reassuring 
check of self--consistency. Note that Eq. \req{tau}
derived previously, exactly reduces to Eq. \req{tau2}
with $\lambda \ll 1$. For $\tau<3$ the second derivative of
$Z(\mu)$ does not exist at $\mu=0$. The second term 
in Eq. \req{tau2} changes, but the leading term remains 
the same:
\[
\tau\cong 2+\frac{2\lambda}{D}-
\frac{2\lambda/D+1}{\Gamma(2\lambda/D+1)}
\int_0^\infty \!dx\frac{e^{-x}-1+x} 
{x^{2+2\lambda/D}}
\left[\frac{4\lambda}{D N}\right]^{\frac{2\lambda}{D}}.
\]

The average growth rate of the capital $N\overline W$ is
obtained summing Eq. \req{dync}
over $i$ and dividing by $N$:
\[
\partial_t \overline W(t)=\left(v+D/2+
\overline{s\tilde{\eta}}\right)\overline W.
\]
The solution to this equation 
\beqar
\overline W(t)&=&\overline W(0)\exp\left[(v+D/2)t+\int_0^t  
\overline{s\tilde{\eta}}(t') dt'\right] \nonumber \\
&\cong& \overline W(0)
e^{
(v+D/2)t
} \nonumber
\eeqar
implies that the growth rate of the average is, to 
leading
order in $N$, equal to the growth rate of the average 
$v+D/2$.

\subsection{Parallels to directed polymers in random media}
In conclusion we would like to point out that 
the stochastic differential equation \req{dync} has a
finite-dimensional analogue, which was much studied 
over the past decade. Indeed, the term $\lambda (\overline W-W_i)$
is nothing else but a fully connected (infinite dimensional) variant
of discrete Laplacian. In finite dimensions this term becomes
$\lambda \Delta W_i=\lambda (\sum_{nn} W_{nn}/2d - W_i)$. 
In the spatial continuous limit the Eq. \req{dync} becomes
\beqar
\partial_t W(x,t)&=&\lambda \Delta W(x,t) + (v+D/2) \ W(x,t)+ \nonumber \\
&+&\eta (x,t) W(x,t), 
\label{polym}
\eeqar
which can be easily recognized as the equation for the partition
function of directed polymer in random media \cite{hz}.
The change of variables $h= \ln W$ maps this equation to the so-called 
KPZ equation \cite{kpz}:
\beq
\partial_t h(x,t)=\lambda (\Delta h(x,t) + |\nabla h (x,t)|^2)+ 
v h(x,t)+\eta (x,t)
\label{KPZ}
\eeq
In our 
infinite-dimensional (fully connected) model
we found that $P(W)$ has a power
law behavior for large $W$. In finite dimensions,
at least below the upper critical dimension $d_c$
(whose very existence is still under debate), this
seems not to be the case. Indeed numerical simulations
show that, at least up to $d=3+1$\cite{kmb91a}
the distribution of $h=\ln W $ has not a pure exponential, 
but rather stretched exponential
behavior. We conjecture that the power law behavior of $P(W)$
in the model studied in this manuscript
is an artifact of the peculiar long range interaction, 
where each site is coupled to any other site.

The work at Brookhaven National Laboratory was 
supported by the U.S. Department of Energy Division
of Material Science, under contract DE-AC02-76CH00016.
S.M. and thanks the Institut de Physique Th\'eorique 
for the hospitality during the visit, when this work was 
initiated. This work is supported in part by the Swiss National
Foundation through the Grant 20-46918.96.

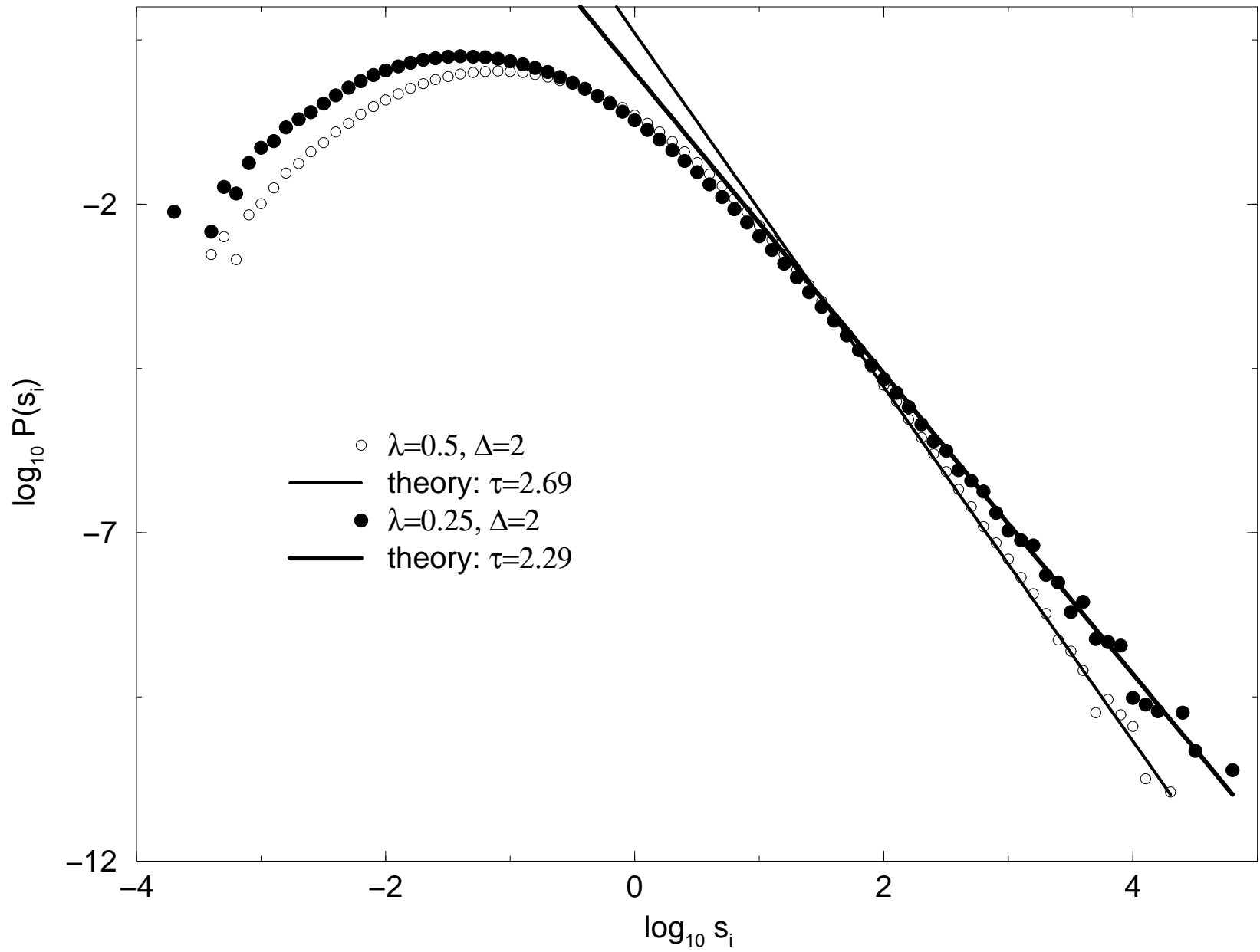
\begin{figure}
\caption{The distribution of capital fractions 
$s_i=W_i/\overline{W}$ for $\lambda=0.25, 0.5$, 
and gaussian $\pi(\eta)$ with $D=2$, and $v=0$ 
in a system of size $N=10000$. The solid lines are the theoretical
predictions (20) for a power law exponent $\tau$ of the 
tail of this distribution.}
\end{figure}

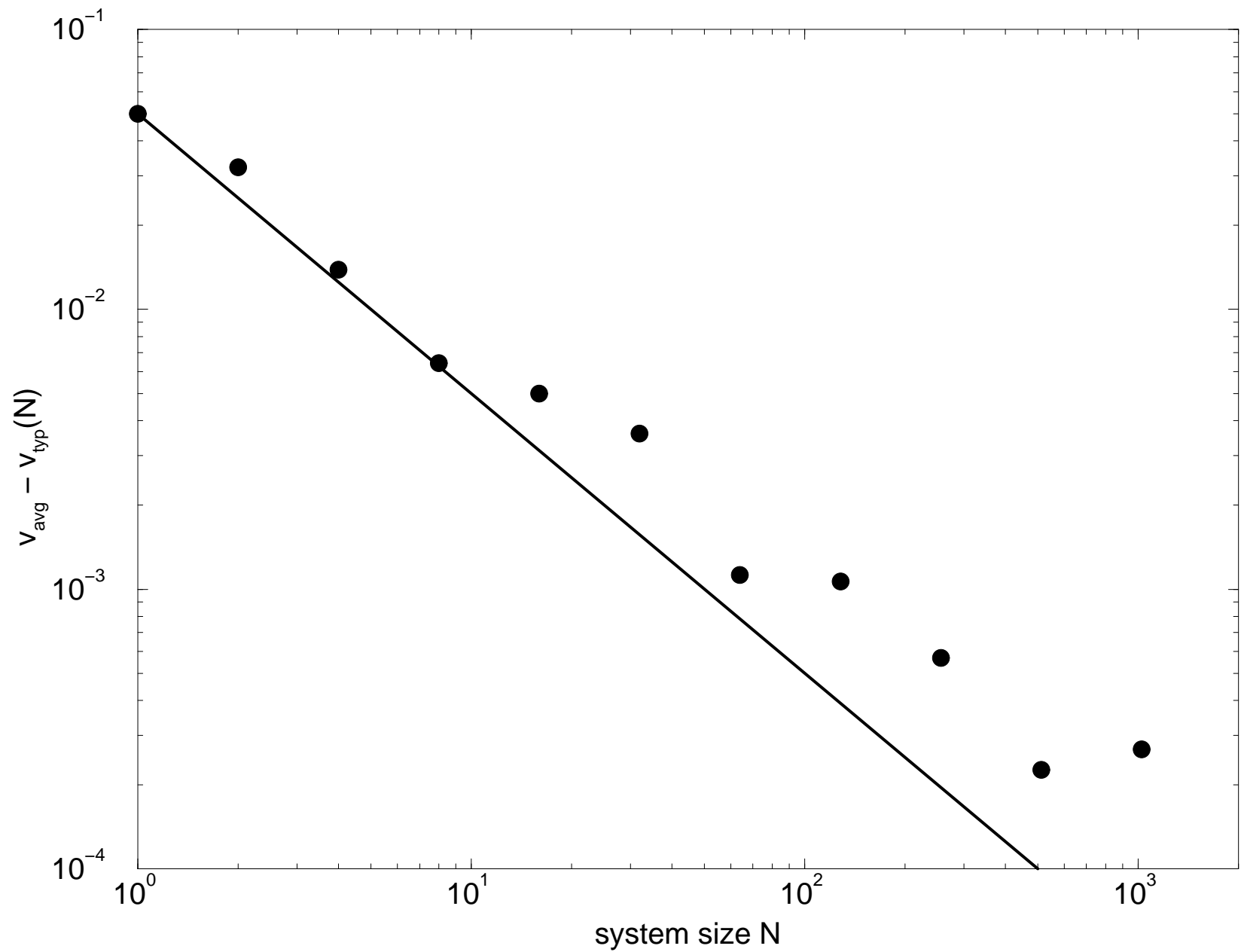
\begin{figure}
\caption{The difference between the average growth rate of the capital
$v_{avg}=D/2$ and its typical growth rate $v_{typ} (N)$ as a function
of the number of assets $N$. The parameters of the model 
are $\lambda=0.1$, $D=0.1$, $v=0$. The solid line indicates 
the theoretical prediction $A/N$. The crossover 
towards smaller $\alpha$ is clearly seen for large $N$.}
\end{figure}

\end{document}